\documentclass[conference]{IEEEtran}
\IEEEoverridecommandlockouts
\usepackage{cite}
\usepackage{amsmath,amssymb,amsfonts,amsthm}
\usepackage{float}
\usepackage{algorithmic}
\usepackage{graphicx}
\usepackage{textcomp}
\usepackage{xcolor}
\usepackage{mathtools}
\usepackage{subcaption}
\usepackage[font=small,labelfont=bf]{caption}
\usepackage{verbatim}

\newcommand{\vX}{\mathbf{X}}
\newcommand{\eX}{{X}}

\newcommand{\vZ}{\mathbf{Z}}
\newcommand{\eZ}{Z}

\newcommand{\vtx}{\mathbf{x}}
\newcommand{\etx}{x}

\newcommand{\vrx}{\mathbf{y}}
\newcommand{\erx}{y}

\newcommand{\ts}{t_\mathrm{s}} 
\newcommand{\Ts}{T_\mathrm{s}} 
   \newcommand{\ths}{\mathrm{th}}
   \newcommand{\vY}{\mathbf{Y}}
   \newcommand{\Si}{\mathrm{S}}
   \newcommand{\Ii}{\mathrm{I}}
   
\newcommand{\Binom}{\mathsf{Binom}}

\newcommand{\Pois}{\mathsf{Pois}}
\newcommand{\prob}[1]{\mathbb{P}\left[#1\right]}

\newcommand{\expect}[1]{\mathbb{E}\left[#1\right]}
\newcommand{\trsp}{\mathsf{T}}

\newtheorem{lemma}{Lemma}

\newtheorem{theorem}{Theorem}

\begin{document}
\addtolength{\topmargin}{.15in}
\addtolength{\textheight}{-.3in}

\title{Optimal Transmission Sequence Design with ISI Matching in Molecular Communication
\thanks{
Email ID: {maravind21, gkrabhi@iitk.ac.in}. The research is supported by Qualcomm and DST SERB grants CRG/23/5206 and MTR/22/390.}
}

\author{Aravind Muraleedharan and Abhishek K. Gupta\\
\IEEEauthorblockA{{Department of Electrical Engineering, Indian Institute of Technology Kanpur}\vspace{0in}
}}

\maketitle

\begin{abstract}
Molecular communication (MC) offers a groundbreaking approach to communication inspired by biological signaling. It is particularly suited for environments where traditional electromagnetic methods fail, such as fluid mediums or within the human body. This study focuses on addressing a major challenge in MC systems: inter symbol interference (ISI), which arises due to the random, diffusive propagation of molecules. We propose a novel technique that leverages transmission shaping to mitigate ISI effectively by designing optimal transmission pulse (or sequence) for symbols. Our approach centers on solving a multi-objective optimization problem that aims to maximize the separability of individual symbol's responses within the symbol duration while matching the interference caused by molecular spillover for all symbols. By making ISI of each symbol similar, the approach reduces the effect of previous symbols and thus not require any adaptive computations. We introduce a geometric analogy involving two families of ellipses to derive the optimal solution. Analytical insights are supported by numerical simulations to design optimized transmission profiles to enhance the resilience toward ISI. The proposed transmission shaping method is evaluated through symbol error rate (SER). These results mark a significant step forward in developing robust and efficient MC systems, opening doors to advanced applications in bio-inspired and nano-scale communication technologies.
\end{abstract}


\section{Introduction}
Molecular Communication (MC) is a bio-inspired paradigm that leverages molecules as information carriers, enabling communication where electromagnetic (EM) waves fail, such as within the human body or underwater. MC operates using bio-nano-machines that transmit and receive information molecules (IMs) through mechanisms like diffusion or active transport \cite{nakano2013}. MC mirrors communications in natural systems, such as cellular signaling in humans or pheromone communication among insects \cite{raguso2008wake}. MC is particularly promising for biomedical applications, including targeted drug delivery and physiological monitoring, due to its bio-compatibility, minimal energy requirements, and nanoscale precision \cite{hiyama2010molecular}.
Encoding schemes such as on-off keying (OOK) and concentration shift keying (CSK) allow the digital transmission of information via molecules termed as information molecules (IMs). In particular, OOK methods involve sending a constant stream of molecules during bit $1$ and none for bit $0$. Despite its potential, MC faces several challenges. The stochastic nature of molecular diffusion introduces latency and uncertainty, often leading to inter symbol interference (ISI) when residual molecules from previous transmissions interfere with the new signal \cite{proakis2008}. This issue is especially pronounced in OOK and CSK systems, where overlapping molecule arrivals compromise data integrity \cite{Nakanonanomachine}. 
To mitigate ISI, researchers have explored diverse strategies, including advanced receptor designs, and error correction schemes tailored to the stochastic nature of MC \cite{ISIavoid}. 
\cite{YilmazISI} analyzed the effect of an adaptive equalizer at the receiver for few modulation schemes. A pre-equalization approach utilizing two types of molecules was proposed in \cite{burcupreeq} to reduce ISI. These methods require adaptive design of pulses based on previously transmitted symbols adding complexity to transmitters and receivers. 
%
%
Authors in  \cite{wang2014transmit} introduced the idea of shaping the molecular transmission profile, suggesting impulse-like or exponentially decaying profiles to reduce the effect of the channel in the received signal. Practical constraints, such as the inability to produce true impulse profiles, led researchers to explore alternative shapes \cite{dhayabaran2021transmit}. Recent experimental efforts, such as \cite{kahvazi2023microfluidic}, have validated these shaping concepts in laboratory conditions, showing promise for real-world applications. \cite{mitrashaping} presented  energy-efficient shaping profiles while \cite{wickeshaping} proposed the use of multiple particle sizes for shaping. 

This paper proposes to utilize the pulse shaping to design a novel optimal transmission pulse (or sequence) for symbols to mitigate the effect of ISI. The approach aims to maximize the
separability of individual symbol's responses within the symbol slot
while matching the ISI of each symbol in subsequent slots. By making ISI of each symbol similar, the approach reduces the effect of previous symbols and thus not requiring any adaptive computations. Analytical insights are supported by numerical simulations via performance evaluation in terms of symbol error. 
By leveraging geometrical analogies and formulating a multi-objective optimization problem, this study proposes shaping profiles that maximize separability while minimizing ISI, paving the way for more reliable MC systems.

\section{System Model}

We consider a 3D MC system consisting of a single transmitter-receiver pair with slotted communication. 
We assume the transmission duration is \(\Ts \) for each symbol. Each transmission  slot is divided into \( N \) sub-slots, with sampling times \( \ts = n\Ts/N \), for \( n = 1, 2, \ldots, N \). Let \( x_{n} \) denotes the number of molecules released in the \( n{\ths} \) sub-time slot corresponding to bit $b$. Let the transmit symbol is denoted as
\begin{equation}
    \vX_b = [\eX_{b1}, \eX_{b2}, \ldots, \eX_{bN}]^T, \quad b \in \{0,1\},
\end{equation}
which is a $n$ dimensional vector. 
We consider a concentration based modulation where for bit 1, the transmitter releases IMs according to the transmission profile vector $\vX_1$, while for bit 0, molecules are released according to the vector $\vX_0$. Let $b[\ell]$ denote the bit value corresponding to $\ell$th slot with our focus on the 0th slot as current slot. Hence, the overall transmit signal is $\vtx$ with $(\ell N+j)$th element equal to $\etx_{N\ell+j}=\eX_{b[\ell],j}$ for $j\in\{1,\cdots,N\}$.



The molecules propagate via the medium to reach the receiver. 
Let $h_k$ denote the probability of an IM being observed at the receiver in the $k$th slot after transmission.
The molecular channel can be modeled as a linear time invariant (LTI) system with a memory of \( M \) symbol slots of durations (\( M\Ts \)). Here, $M$ denotes the maximum slot index up to which the IMs can be observed from a transmission.
Hence, the channel impulse response is sampled over \( (M+1)N \) time points, forming the sequence: 
Hence, the channel response is given as \([h_1, h_2, h_3, \ldots, h_{(M+1)N}].\) which spans the current symbol duration \( T_s \) and \( M \) future slots, accounting for ISI caused by the diffusive propagation of molecules.

Note that the meaning of observed molecules depends on the receiver type. For a receiver with fully absorbing spherical surface that captures and counts the arriving molecules, $h_k=\int_{(k-1)\ts}^{k\ts} \kappa(t)\mathrm{d}t$ where $\kappa(t)$  denotes the hitting rate given as
\begin{equation}
    \kappa(t) = \frac{a}{R+a} \frac{R}{\sqrt{4 \pi D t^3}} e^{-\frac{R^2}{4Dt}},
\end{equation}
where \( D \) is the diffusion coefficient of IMs, \( R \) is the transmitter-receiver distance, and \( a \) is the receiver radius. For a receiver with passive receiver with small volume $v$ that counts the  molecules inside itself at any time instant, $h_k=h(k\ts)$ with
\begin{equation}
    h(t)= \frac{v}{\sqrt{(4 \pi D t)^3}} e^{-\frac{R^2}{4Dt}}.
\end{equation}
Although the presented analysis is true for a wide range of receivers, we consider a passive receiver in the rest of the paper for brevity.

\section{Receive Signal and  Decoder}
Let vector $\vrx=[y_1,\cdots, y_N]$  denote the observed molecules in the 0th slot a with $y_k$ denoting the molecule observed at the receiver at $k$th time subslot (i.e between $(k-1)\ts$ and $k\ts$).
The contribution $y_k^{(x_j)}$ in received signal $y_k$ due to transmission $x_j$ where $j\in[k-(M+1)N+1\cdots k]$ is a random variable (RV) with binomial distribution $\Binom(x_j,h_{k-j+1})$. Due to passive receivers, $y_k^{(x_j)}$ are independent  over $j$ and $k$.  Hence, $y_k$ is sum of $(M+1)N$ binomial random variable plus an additional $n_k$ denoting noise with Poisson distribution i.e. $n_k\sim\Pois(\lambda)$. Therefore,
\begin{align}
y_k=\sum\nolimits_{j=1}^{(M+1)N} y_k^{(x_j)}+n_k.\label{eq:y_main}
\end{align}
Hence, the PMF of $y_k$ is given as
\begin{align*}
&\prob{y_k=m\,|\,b[\ell],l\in[-M,\cdots 0]}\\
&=\sum_{\sum_j i_j +i_\mathrm{n}=m}  \prod_{j} \binom{x_{k-j+1}}{i_1} h_{j}^{i_j}(1-h_j)^{x_{k-j+1}-i_j}\frac{\lambda^{i_\mathrm{n}}}{i_\mathrm{n}}e^{-\lambda}
\end{align*}
where $x_j$'s depend on the current and previous symbols $\vX_{b[\ell]}$ for $\ell\in\{-M,-1,\cdots,0\}$.
Utilizing $\expect{y_k^{(x_j)}}=x_jh_{k-j+1}$ in \eqref{eq:y_main}, the mean of $\mathbf{y}$ is given as 
\begin{align*}
&\mu_b\stackrel{\Delta}{=}\expect{\mathbf{y}|b[0]=b,\{b[\ell]\}} =\underbrace{\vphantom{\sum\nolimits_d}H_\Si \vX_b}_{\vY_{b\Si}} + \underbrace{\sum\nolimits_{\ell=1}^{M} H_{\Ii_\ell} \vX_{b[t-\ell]} }_{\vY_{\Ii}}+ \lambda \mathbf{1} 
\\
& \text{where }   H_\Si = \begin{bmatrix}
\quad \ h_1 \quad\ & \ \quad0\quad & \quad0\quad & ... &\quad \ \ \  0\ \ \quad\ \\
h_2 & h_1 & 0 & ... & 0 \\
h_3 & h_2 & h_1 & ... & 0 \\
\vdots & \vdots & \vdots & \vdots & \vdots \\
h_N & h_{N-1} & ... & h_2 & h_1 \\
\end{bmatrix},\\
&\qquad \ \ H_{\Ii_i}= \begin{bmatrix}
h_{iN+1} & h_{iN} & ... & h_3 & h_2 \\
\vdots & \ddots & ... & \vdots & \vdots \\
h_{(i+1)N} & h_{(i+1)N-1} & ... & h_{iN+2} & h_{iN+1} \\
\end{bmatrix},
\end{align*}
 and $\mathbf{1}$ is all ones $N\times 1$ column. Here, the $H_\Si$ and all $H_{\Ii_i}$'s are of dimensions $N \times N$. Although we consider that transmit sequence and received signal has the sampling frequency, it is trivial to extend to the case where they are different, leading to $H$'s being rectangular matrices instead.

An optimal ML decoding rule is to decode 1 when $\vrx\in D(1)$ with 
\begin{align*}
D(1)=\{p(\mathbf{y}|b[0]=1)>p(\mathbf{y}|b[0]=0)\}
\end{align*}
where
\begin{align*}
p(\mathbf{y}|b[0]=b)=\expect{\prod_{k=1}^N \prob{y_k=m|b[\ell] } |b[0]=b }
\end{align*}
denotes the likelihood of $\vrx$ for two values of bit. 
The bit error probability (BER) is given as 
\begin{align}
p_e= \frac12\left(\sum_{\vrx\in D(1)} p(\vrx|b[0]=0)+\!\!\!\!\sum_{\mathbf{y}\notin D(1)} p(\vrx|b[0]=1)\right). \label{eq:ber_main}
\end{align}

\section{Optimal Design of Transmit Sequence}

The optimal transmit sequence can be obtained by minimizing $p_e$ as given in \eqref{eq:ber_main} over all possible values of $\vX_0$ and $\vX_1$, in principle. However, the optimization is difficult 
owing the complexity of $p_e$ expression. Hence, we present a method based on statistical approximation. Note that the BER  depends on statistical difference between $\mathbf{y}$ for $b[0]=0$ and $1$, averaged over all possible values of previous bits $b[\ell]$'s. 

To ensure statistical separability between $\vrx$ for bit 0 and 1, it requires a large difference between $\mu_0$ and $\mu_1$ which can be obtained by creating a significant difference between signal terms $\vY_{0\Si}=H_\Si X_0$ and $\vY_{1\Si}=H_\Si X_1$. 
However, despite the large separation between $\mu_0$ and $\mu_1$, due to previous bits, the ISI term $\vY_{\Ii}$ can move this separation which makes it difficult to utilize this separation to design decoding region. One method is to change decoding regions adaptively based on previous bits which is difficult to implement. Another approach is to scale $\vX$ to increase the separation further does not work as $\vX$ will increase  the effect of previous bits. 
 
In this paper, we take novel approach where we try to match the ISI terms $\vY_{0\Ii_\ell}=H_{\Ii_\ell}\vX_0$ and $\vY_{1\Ii_\ell}=H_{\Ii_\ell}\vX_1$ for all $i$ to ensure that the ISI term $\vY_{\Ii}$ does not change depending on previous bits.
Let vector  $\vY_{b\Ii}$ denote the observed molecules in subsequent $M$ slots for bit 0 and is given as
\begin{align*}
\vY_{b\Ii}
 = \begin{bmatrix}
     \vY_{b\Ii_1} \\ \vY_{b\Ii_2} \\ \vdots \\ \vY_{b\Ii_M}
\end{bmatrix}=H_\Ii \vX_b, 
  \text{ where }
     H_\Ii
 = \begin{bmatrix}
     H_{\Ii_1} \\ H_{\Ii_2} \\ \vdots \\ H_{\Ii_M}
\end{bmatrix}.
\end{align*}
The  goal is to design \( \vX_1 \) and \( \vX_0 \) such that the difference in the ISI $\vY_{b\Ii}$ for two bit values is minimized while ensuring separability of signal terms $\vY_{b\Si}$  for them. 
On a general note, subscript $\Si$ denotes the symbol (information for that particular slot) component and subscript $\Ii$ denotes the interfering (spreading) (ISI) component.

Our objective is to optimize the molecule release profiles \({X}_1 \) and \({X}_0\) (also called transmission sequences) such that:
\begin{itemize}
    \item $\vY_{1\Si} = H_\Si \vX_1$ and $\vY_{0\Si} = H_\Si \vX_0$ are maximally separable, indicating that they are distinct responses and carry meaningful information relevant to the intended symbol.
    \item $\vY_{1\Ii} = H_\Ii \vX_1$ and $\vY_{0\Ii} = H_\Ii \vX_0$ are maximally similar, to result in similar interference across future slots.
\end{itemize}
Note that our approach to modeling ISI differs from most existing works, where ISI is typically characterized as interference caused by a symbol overlapping with previously transmitted symbols. In contrast, here, we focus on analyzing the impact of a transmitted symbol on subsequent future symbols. 

\subsection{Optimization Problem}
This equivalent optimization problem can be formulated as
\begin{align*}
         \max_{\vX_1, \vX_0} & \| H_\Si (\vX_1 - \vX_0)\|^2, \\
         \min_{\vX_1, \vX_0} & \| H_\Ii (\vX_1 - \vX_0) \|^2, \\
         \text{s.t. } &\vX_1, \vX_0 \ge 0.
\end{align*}
Here, the constraint $\vX_1, \vX_0 \ge 0$ implies that all the individual elements of the vectors $\vX_1$ and $\vX_0$ are non-negative. Further, there can be additional constraints, for example,  $\vX_b^\trsp\mathbf{1}\le P$ which puts a limit $P$ on the total number of IMs released for each bit value, which we will discuss in next section.
 
Let us define transmission difference vector $\vZ=\vX_1-\vX_0$, $A=H_\Si^\trsp H_\Si$, and $B=H_\Ii^\trsp H_\Ii$. Hence, we have
\begin{align}
\|H_\Si(\vX_{1}-\vX_{0})\|^2 &= (\vX_1-\vX_0)^\trsp H_\Si^\trsp H_\Si(\vX_1-\vX_0) \nonumber
\\ &= (\vX_1-\vX_0)^\trsp A(\vX_1-\vX_0) = \vZ^\trsp A\vZ,\nonumber
\end{align}
and
\begin{align}
\|H_\Ii(\vX_{1}-\vX_{0})\|^2  &= (\vX_1-\vX_0)^\trsp H_\Ii^\trsp H_\Ii (\vX_1-\vX_0) \nonumber\\
& = (\vX_1-\vX_0)^\trsp B(\vX_1-\vX_0) = \vZ^\trsp B\vZ.
\end{align}

Hence, an equivalent optimization problem is given as
    \begin{align}
    \mathsf{P: \ }    & \max_{\vZ} \; \vZ^\trsp A\vZ, \ 
        & \min_{\vZ} \; \vZ^\trsp B\vZ,
    \end{align}
which is a multi-objective optimization. We can turn this problem into a single-objective optimization problem by turning one of the objective into constraint.

\textbf{Problem 1}: If we put a constraint on the separation between $\vY_{1\Si}$ and $\vY_{0\Si}$ be at least $\epsilon_0$, to ensure a certain statistical distinguishability, then the optimization problem turns into minimizing the ISI difference $\|\vY_{1\Ii}-\vY_{0\Ii} \|^2 = \vZ^\trsp B\vZ$, i.e. 
\begin{align}
\mathsf{P1:\ }  \min_{\vZ} \ &\vZ^\trsp B \vZ\\
   \text{s.t. } &  \vZ^\trsp A\vZ \geq \epsilon_0.
\end{align}

\textbf{Problem 2}: If we put a constraint on the difference between $\vY_{1\Ii}$ and $\vY_{0\Ii}$ to be at max $l_0$, to limit the effect of previous bits, then the optimization problem turns  into maximizing signal separation $\|\vY_{1\Si}-\vY_{0\Si} \|^2 = \vZ^\trsp A\vZ$, 
\begin{align}
\mathsf{P2:\ }  \max_{\vZ} \ &\vZ^\trsp A \vZ\\
   \text{s.t. } &  \vZ^\trsp B\vZ \le l_0.
\end{align}

To create a better exposition, we will consider an example consisting a special case of $N=2$ first and demonstrate the optimization methodology. 

\subsection{Example: Case $N=2$.}
Here, for $N=2$, we have
\begin{align*}
&\vX_0 = \begin{bmatrix}
    \eX_{01} \\
    \eX_{02}
\end{bmatrix}, 
&&\vX_1=
\begin{bmatrix}
     \eX_{11} \\
     \eX_{12}
\end{bmatrix}, \qquad 
\vZ=
\begin{bmatrix}
     \eZ_{1} \\
     \eZ_{2}
\end{bmatrix}\\
\text{and }
& H_\Si=\begin{bmatrix}
    h_1 & 0 \\
    h_2 & h_1
\end{bmatrix}, 
&&H_\Ii=\begin{bmatrix}
    h_3 & h_2 \\
    h_4 & h_3
\end{bmatrix}.
\end{align*}

Here, for $\mathsf{P}1$, it can be seen that the constraint
\begin{align}
 \vZ^\trsp A \vZ = (\vX_1-\vX_0)^\trsp A (\vX_1-\vX_0) \ge \epsilon_0\nonumber
 \end{align}
 
is equivalent to
\begin{align*}
\begin{bmatrix}
    \eX_{11}-\eX_{01} & \eX_{12}-\eX_{02}
\end{bmatrix}
\begin{bmatrix}
    h_1 & h_2 \\
    0 & h_1
\end{bmatrix} 
\begin{bmatrix}
    h_1 & 0 \\
    h_2 & h_1
\end{bmatrix}
\begin{bmatrix}
    \eX_{11}-\eX_{01} \\
    \eX_{12}-\eX_{02}
\end{bmatrix} &\\
\ge \epsilon_0&\\
    (h_1^2+h_2^2)(\eX_{11}-\eX_{01})^2 + 2h_1h_2(\eX_{12}-\eX_{02})(\eX_{11}-\eX_{01})\quad &\\+ h_1^2(\eX_{12}-\eX_{02})^2 \ge \epsilon_0&,\\
(h_1^2+h_2^2)(\eZ_1)^2 + 2h_1h_2(\eZ_1\eZ_2) + h_1^2(\eZ_{2})^2 \ge \epsilon_0&,
\end{align*}
which represents an ellipse in the $\eZ_1$-$\eZ_2$ plane, where $\eZ_1 = \eX_{11}-\eX_{01}$ and $\eZ_2=\eX_{12}-\eX_{02}$.
Similarly, it can also be seen that \(\vZ^\trsp B\vZ=l_0\) is another ellipse in the $\eZ_1$-$\eZ_2$ plane.

\begin{figure}[H]
\centering
        \includegraphics[width=.47\linewidth, trim=0 0 0 0, clip]{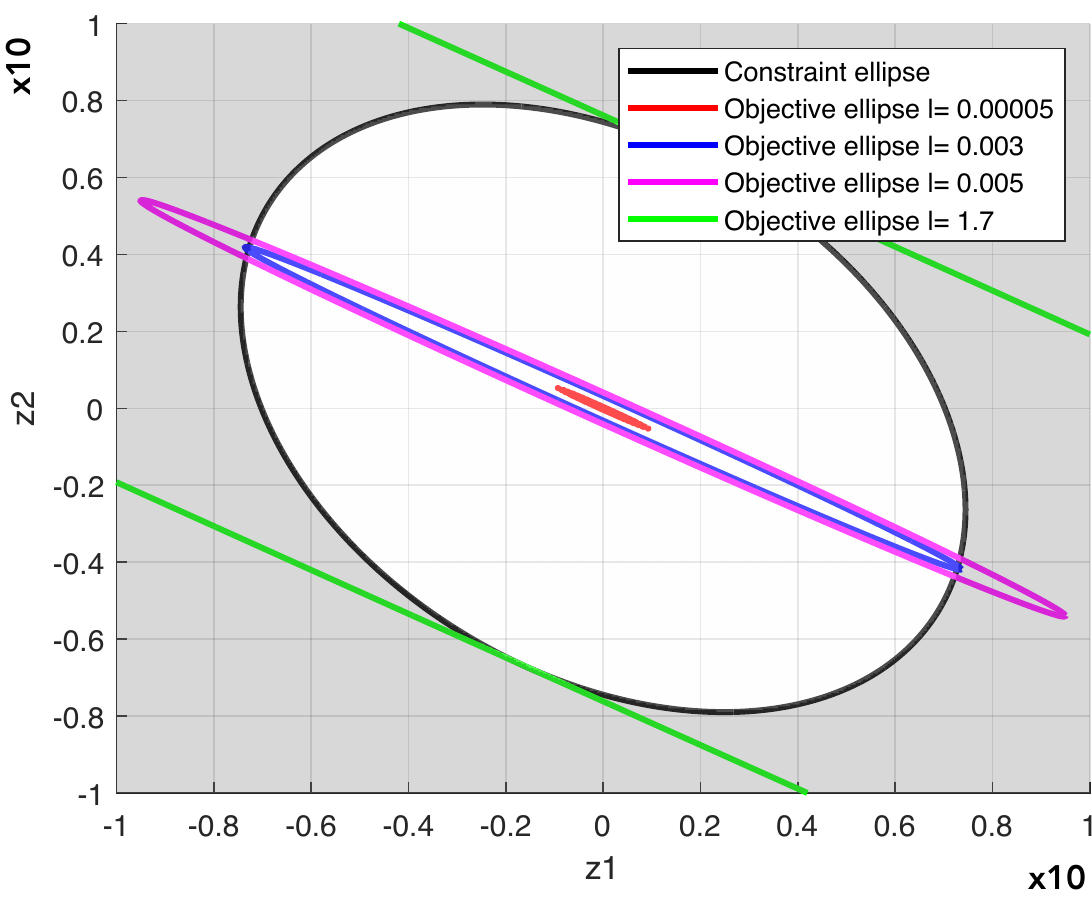}
        \includegraphics[width=.47\linewidth]{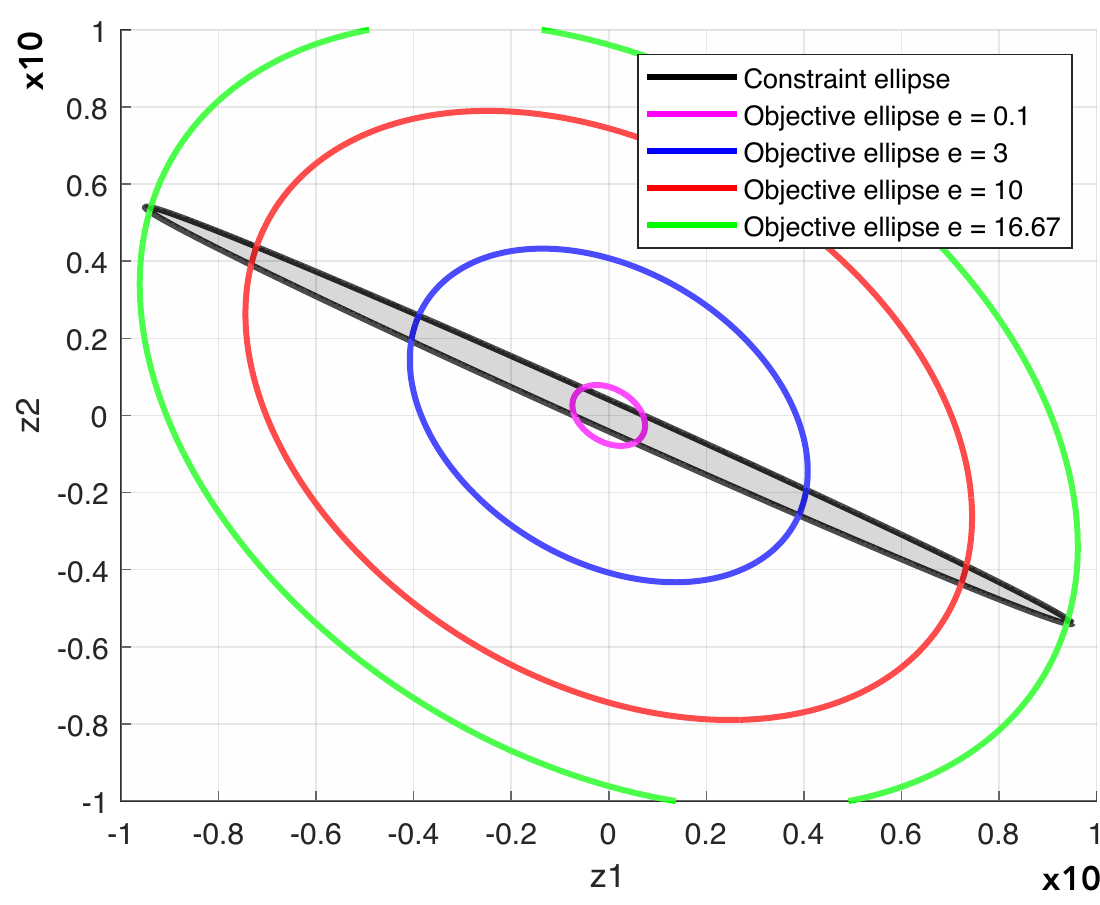}
\\ {\small \bf (a)\hspace{2in} (b)}
\caption{
Geometric interpretation of the optimization for the case $N=2$ with $h=[h_1\, h_2\, h_3\, h_4]=[4.24\, 1.5\, 0.81\, 0.53]\times 10^{-2}$ for (a) $\mathsf{P1}$ with constraint ellipse $\vZ^\trsp A\vZ \ge \epsilon_0=10$ and objective ellipses denoted by  $\vZ^\trsp B\vZ=l$,  for different values of $l$  
%
and (b) $\mathsf{P2}$ with constraint ellipse $Z^TBZ \leq l_0=0.005$ and objective ellipses denoted by  $Z^TAZ=\epsilon$ for different values of $\epsilon$. 
}
\label{fig:1}
\end{figure}

It can be seen that the optimization process in each of the two problems is centered on the interaction between an objective ellipse and a constraint ellipse in the \( \vZ \)-space.
In {\emph{Problem 1}}, the goal is to minimize the objective ellipse such that it intersects with the region outside the constraint ellipse.  To minimize \( \vZ^\trsp B \vZ \), we evaluate a family of ellipses represented as \( \vZ^\trsp B \vZ = l \), for varying \( l \). The ellipse with smallest $l$ that intersects the exterior region spanned by \( \vZ^\trsp A \vZ \ge \epsilon_0 \) gives us the optimal $\vZ$.  Hence, the optimum occurs when the objective ellipse touches the constraint ellipse. Fig. \ref{fig:1}(a) provide a 2D geometric interpretation of these approach.

Similarly,  in {\emph{Problem 2}}, the objective is to maximize the objective ellipse while it remains inside the constraint ellipse. To maximize \(\vZ^\trsp A \vZ \), we consider the family \( \vZ^\trsp A \vZ = \epsilon \), for varying \( \epsilon \). The biggest ellipse that lies inside the interior region spanned by \( \vZ^\trsp B \vZ \leq l_0 \) gives us the optimal $Z$. 

In both cases, the optimal solution corresponds to the points where  the objective and constraint ellipses touch each other. For higher dimensions, these become ellipsoids. 

We can see that each ellipse reduces to a pair of parallel lines  when $H_S$ or $H_I$ are not
of full rank.  Further, ellipses represent $N$ dimensional
hyper-ellipsoids for higher $N$. This means they represent pair of points for $N=1$.

\subsection{Optimal Transmission Difference Vector for General Case}
We present the following three lemmas to aid in an analytical approach for a general \( N \) from \cite{boydconvex}.

\begin{lemma}\label{lem:1}
Consider the optimization
\[
\text{minimize} \quad w^\trsp R w \quad \text{s.t.} \quad w^\trsp w = 1,
\]
where $R$ is positive semi-definite. The optimal solution $w$ is the normalized eigenvector corresponding to the smallest eigenvalue of the matrix \( R \), and the minimum value of the objective function is the smallest eigenvalue of \( R \) \cite{boydconvex}.
\end{lemma}

\begin{IEEEproof}
See Appendix \ref{app:a}.
\end{IEEEproof}

\begin{figure}[H]
\centering
        \includegraphics[width=.47\linewidth]{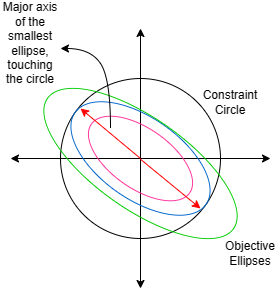}
        \includegraphics[width=.47\linewidth]{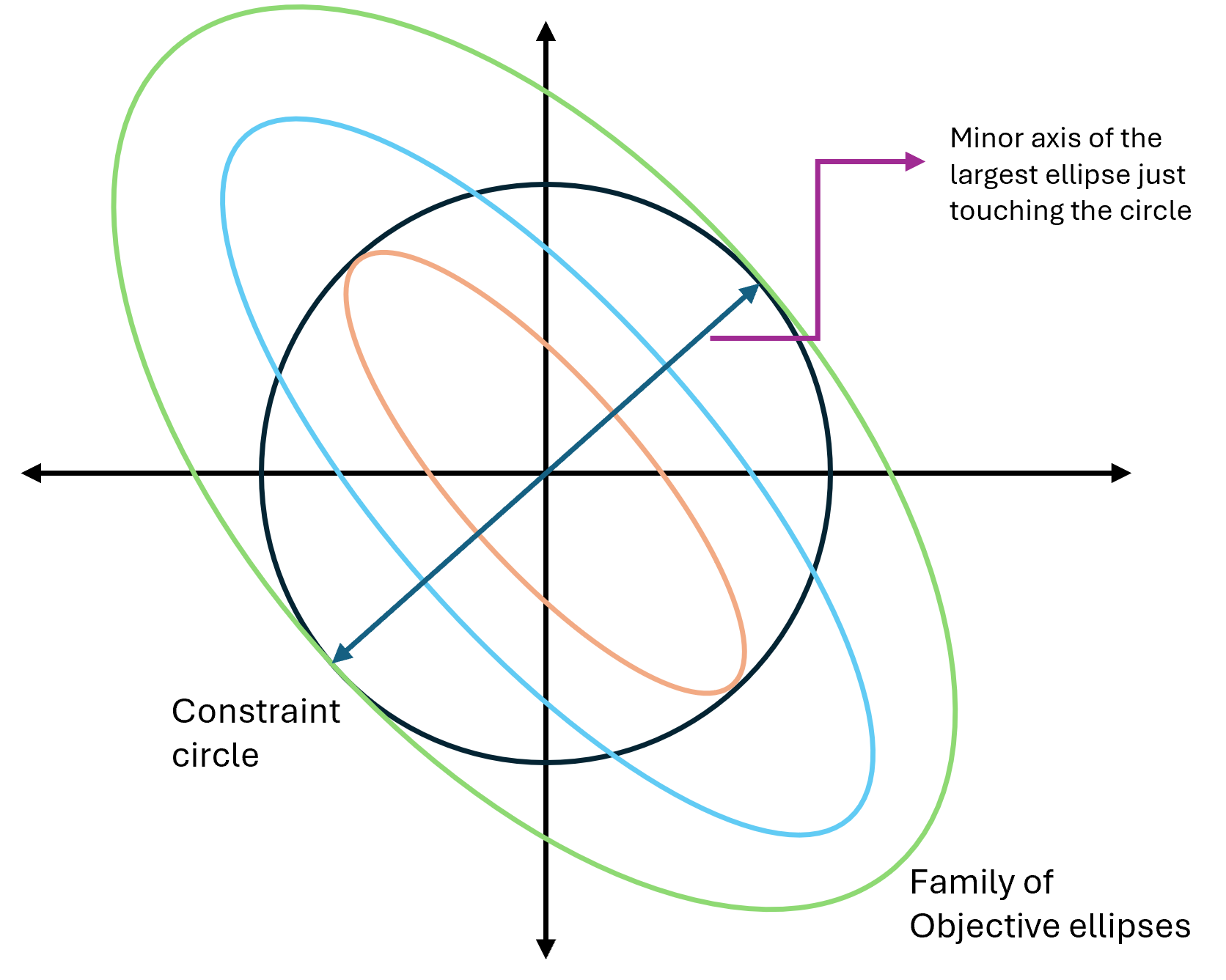}
\\ {\small \bf (a)\hspace{2in} (b)}
\caption{The geometric interpretation of the optimization in (a) Lemma 1 and (b) Lemma 2.}
\label{fig:2}
\end{figure}

Geometrically, this problem is equivalent to finding the point on the unit circle \( w^T w = 1 \) that lies on the smallest ellipse \( w^T R w = q \). (See Fig. \ref{fig:2}(a) for an illustration). The smallest ellipse intersects the circle along its major axis, corresponding to the eigenvector of the smallest eigenvalue. 

For a family of ellipses, $w^TRw=e$, with respect to $e$, the ellipse grows fastest along the minor axis (largest eigenvalue direction). The family of ellipses with varying $e$ is densest along the major axis (smallest eigenvalue direction).


\begin{lemma}\label{lem:2}
For the optimization
\[
\text{maximize} \quad w^\trsp R w \quad \text{s.t.} \quad w^T w = 1,
\]
where $R$ is semi positive definite, the optimal value $w$ is the normalized eigenvector corresponding to the largest eigenvalue of \( R \), and the maximum value of the objective function is the largest eigenvalue of \( R \).
\end{lemma}

\begin{IEEEproof}
See Appendix \ref{app:b}.
\end{IEEEproof}

Geometrically, this corresponds to the unit circle intersecting the largest ellipse along its minor axis (see Fig. \ref{fig:2}(b)).


\begin{lemma}\label{lem:3}
For positive definite matrices \( A \) and \( B \), 
\[
\text{minimize} \quad z^\trsp B z \quad \text{s.t.} \quad z^T A z \geq \epsilon_0,
\]
yields \( z =\sqrt{\epsilon_0}A^{-1/2}w\) with $w$ as the eigenvector corresponding to the smallest eigenvalue $\lambda_{\mathrm{min}}$ of \( R = A^{-\frac{1}{2}} B A^{-\frac{1}{2}} \), with a minimum value \( \epsilon_0 \lambda_{\mathrm{min}} \). Similarly, the problem
\[
\text{maximize} \quad z^T A z \quad \text{s.t.} \quad z^T B z \leq l_0,
\]
yields \( z =\sqrt{l_0}B^{-1/2}u\) with $u$ as the eigenvector corresponding to the largest eigenvalue $\lambda_{\mathrm{max}}$ of \( R' = B^{-\frac{1}{2}} A B^{-\frac{1}{2}} \), with a maximum value \( l_0\lambda_{\mathrm{max}}\).
\end{lemma}

\begin{IEEEproof}
See Appendix \ref{app:c}.
\end{IEEEproof}


Building on these results, we get the following main result as given in the following theorem.

Note that $A^{-1/2}w$  is equivalent to  $B^{-1/2}u$. 
Building on these results, we get the following main result as given in the following theorem.


\begin{theorem}\label{thm:1}
The optimal transmission difference vector $\vZ$ for problem $\mathsf{P}$ is $\beta A^{-1/2}\mathbf{w}$ where $\mathbf{w}$ is the eigen vector corresponding to the smallest eigen value of $A^{-\frac{1}{2}} B A^{-\frac{1}{2}}$ with  $A=H_\Si^\trsp H_\Si$, and $B=H_\Ii^\trsp H_\Ii$. Here, $\beta$ is arbitrary coefficient determining the trade-off between two objectives. 
\end{theorem}

\subsection{Optimal Transmission Sequences}
For a given $\vZ$, there are multiple possibilities of $\vX_0$ and $\vX_1$. Further, constraints or objectives can be put from the perspective of energy efficiency, molecular cost etc. Here, we consider the goal of minimizing molecular cost, where given optimal $\vZ$, the optimal transmission vector solves the following problem
\begin{align*}
\mathsf{M1}: \min &\ \mathbf{1}^\trsp (\vX_0+\vX_1)\\
\text{s.t.} &\ \vX_1-\vX_0=\vZ, \vX_0 \ge 0, \vX_1\ge 0.
\end{align*}
The solution of above problem is given in the following theorem.

\begin{theorem}\label{thm:2}
Given optimal $\vZ$ as obtained in Theorem \ref{thm:1}, the optimal transmission sequences  is given as
\begin{equation}
    \vX_b = [\eX_{b1}, \eX_{b2}, \ldots, \eX_{bN}]^T, \quad b \in \{0,1\},
\end{equation}
\begin{align}
\text{with }
\eX_{1i}=\begin{cases} \eZ_i &\text{if } \eZ_i\ge 0\\
0  &\text{if } \eZ_i<0, 
\end{cases}
&&
\eX_{0i}=
\begin{cases} 0&\text{if } \eZ_i\ge 0\\
-\eZ_i &\text{if } \eZ_i<0\
\end{cases}\nonumber
\end{align}
\end{theorem}

\begin{IEEEproof}
Since all constraints are separable across $\eX_{bj}$, the minimum is achieved by solving the following for each $j$
\begin{align*}
\min &\ \eX_{0j}+\eX_{1j}\\
\text{s.t.} &\ \eX_{1j}-\eX_{0j}=\eZ_j, \eX_{0j} \ge 0, \eX_{1j}\ge 0.
\end{align*}
When $\eZ_j\ge0$, we can write $\eX_{1j}=\eX_{0j}+\eZ_j,$ to get
\begin{align*}
\min &\ 2\eX_{0j}+\eZ_{j}, \quad  \text{s.t. } \eX_{0j} \ge 0.
\end{align*}
Hence, the solution is $\eX_{0j}=0$ and $\eX_{1j}=\eZ_j$. The steps can be repeated for the case $\eZ_j<0$. 
\end{IEEEproof}

\begin{figure}[htbp]
    \centering
      \includegraphics[width=\linewidth,trim=50 0 0 0,clip]{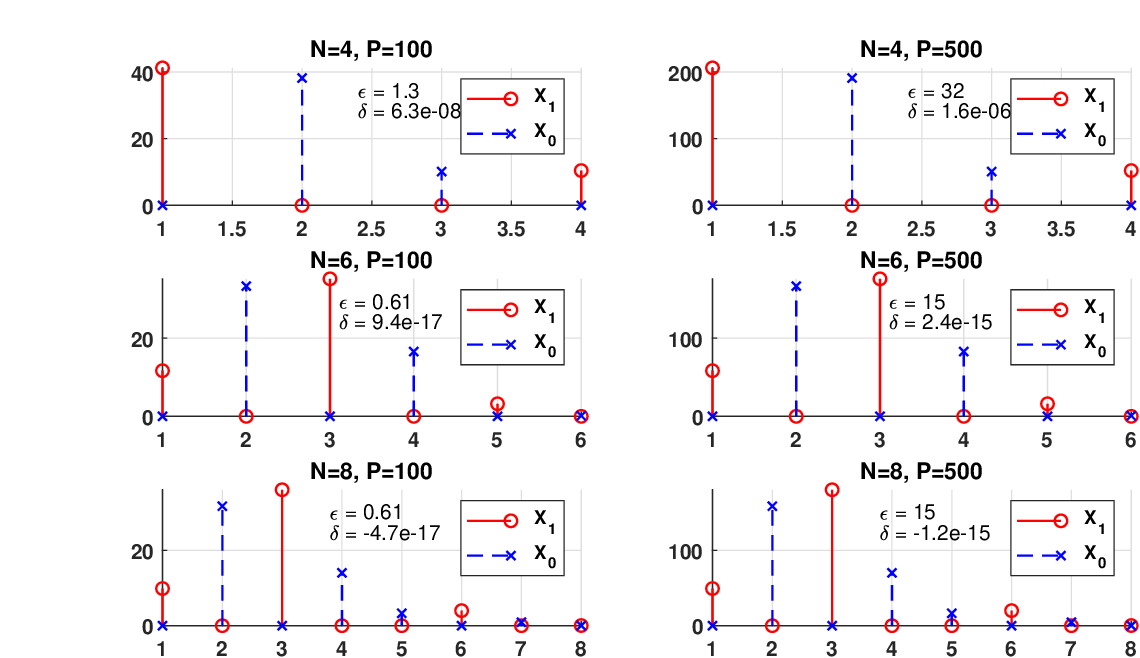}
    \caption{Optimal transmit sequence for bit 0 and 1 for different values of $N$ and $P$ under constraint on total number of IMs released.
    Values of $\epsilon= \vZ^\trsp A\vZ$ and $\delta= \vZ^\trsp B\vZ$ are also displayed. Here, $D=10^{-8}m^2/s$, $R=100\mu m$m and the receiver radius is $30\mu m$.  
    }
    \label{fig:figN_P}
\end{figure}

Examples of other constraints include  $\mathbf{1}^\trsp \vX_b\le P$ which puts a limit $P$ on the total number of IMs released for each bit value or $\mathbf{1}^\trsp (\vX_1+\vX_0)= P$ which fixes the total number of IMs released to be $P$. The solutions of the later problem is $\overline{\vX_b}=(P-\mathbf{1}^\trsp (\vX_0+\vX_1))/(2N)\mathbf{1}+\vX_b$ where $\vX_b$ is given in Theorem \ref{thm:2}, assuming $P$ is large enough. 
Another problem formulation that  fixes the total number of IMs released to be $P$ can be written as
\begin{align*}
\mathsf{M2}:  \max &\ \beta\\
\text{s.t.} &\ \vX_1-\vX_0=\beta\vZ, \mathbf{1}^\trsp (\vX_1+\vX_0)= P,  \vX_0 , \vX_1\ge 0.
\end{align*}
The solution  is $\overline{\vX_b}=\beta\vX_b$ with $\beta=P/(\mathbf{1}^\trsp (\vX_0+\vX_1))$
where $\vX_b$ is given in Theorem \ref{thm:2}.
Fig. \ref{fig:figN_P} shows the shape behavior of $\vX_1$ and $\vX_0$ for varying $N$ and $P$ solved according to $\mathsf{M2}$. 
As $P$ increases, the magnitudes of the sequences increase, but shape is almost preserved. It can be observed that with large $N$, a better tradeoff can be achieved.

In the above discussions, we have assumed that the number of released IMs is large enough to avoid any integer constraints on $\vX_b$. In otherwise cases, the solution can be converted to closest integer solution or additional constraints can be put on $\vX_b$'s.  

%
\section{Performance Analysis}
We now present the performance of the proposed transmission shaping. 
As described earlier, given previous bits, $y_k^{(x_j)}$'s are independent binomial RVs i.e. $y_k^{(x_j)}\sim \Binom(x_j,h_{k-j+1})$. Hence, the distribution of $y_k$ given as
\begin{align}
y_k=\sum\nolimits_{j=1}^{(M+1)N} y_k^{(x_j)}+n_k
\end{align}
is complicated. However, under assumption of large $x_j$ and small $h_j$, binomial can be distributed well by Poisson distribution. Applying the fact that sum of Poisson RVs is also Poisson, $y_k$ turns out to be Poisson RV with mean $ \sum\nolimits_{j=1}^{(M+1)N} x_j h_{k-j+1}+\lambda_k$. Hence, given current and previous bits, $\vrx$ is a Poisson random vector with mean  
\begin{align*}
\mu_b=H_\Si \vX_b+\sum\nolimits_{\ell=1}^{M} H_{\Ii_\ell} \vX_{b[t-\ell]} + \lambda \mathbf{1} 
\end{align*}
with each element independently distributed as Poisson. 

\subsection{Optimal ML decoder}
As stated earlier, the optimal ML decoding rule is to decode 1 when $\vrx\in D(1)$ with 
$D(1)=\{p(\mathbf{y}|b[0]=1)>p(\mathbf{y}|b[0]=0)\}$
where $p(\mathbf{y}|b[0]=b)$ denotes the likelihood of $\vrx$ for two values of bits. Assuming that the ISI term deviates only by small amount based on previous bit $b[\ell]$, we can approximate $\vrx |b[0]$ as Poisson with mean 
\begin{align*}
\mu_b=H_\Si \vX_b+\frac12 \sum\nolimits_{\ell=1}^{M} H_{\Ii_\ell} (\vX_{1}+\vX_{0}) + \lambda \mathbf{1} .
\end{align*}
This means that 
\begin{align}
p(\mathbf{y}|b[0]=b)=\prod_{k=1}^N \exp(-\mu_{bk}) \frac{\mu_{bk}^{\erx_k}}{\erx_k!}\label{eq:y_dist}
\end{align}
where $\mu_{bk}$ is the $k$th element of $\mu_b$. Applying the decoding rule, we get the following result.

\begin{theorem}\label{thm:opt_dec}
Under Poisson channel approximation, the ML
decoding region for bit 1  is
given as
\begin{align}
D(1)=\left\{ \vrx: \sum a_k \erx_k >c
\right\}
\end{align}
with $a_k=\ln\left(\frac{\mu_{1k}}{\mu_{0k}}\right)$  and $c=\mathbf{1}^\trsp 
 (\mu_{1}-\mu_{0})$.
\end{theorem}

\begin{IEEEproof}
Under ML decoding, the bit is decoded as 1 if
\begin{align*}
p(\mathbf{y}|b[0]=1)>p(\mathbf{y}|b[0]=0).
\end{align*}
Using \eqref{eq:y_dist}, the above is equivalent to
\begin{align*}
\prod_{k=1}^N \exp(-\mu_{1k}) \frac{\mu_{1k}^{\erx_k}}{\erx_k!}
>
\prod_{k=1}^N \exp(-\mu_{0k}) \frac{\mu_{0k}^{\erx_k}}{\erx_k!}.
\end{align*}
Applying logarithm on both sides, we have
\begin{align*}
\sum_{k=1}^N -\mu_{1k}+{\erx_k} \ln \mu_{1k} >
\sum_{k=1}^N -\mu_{0k}+{\erx_k} \ln \mu_{0k} .
\end{align*}
Simplifying it, we get the desired result.
\end{IEEEproof}

Note that $ \mu_{1}-\mu_{0}=H_\Si (\vX_1-\vX_0)$. It is important to note that all samples are not weighted equally in the decoder.

\subsection{Bit error probability}

Utilizing \eqref{eq:ber_main}, the BER is given as 
\begin{align}
p_e= \frac12\left(\sum_{\sum a_k y_k>c} \prod_{k=1}^N e^{-\mu_{0k}} \frac{\mu_{0k}^{\erx_k}}{\erx_k!}+\sum_{\sum a_k y_k\le c} 
\prod_{k=1}^N e^{-\mu_{1k}} \frac{\mu_{1k}^{\erx_k}}{\erx_k!}\right). \nonumber
\end{align}

\appendices

\section{Proof of Lemma \ref{lem:1}}\label{app:a}
 The Lagrangian for this problem is given as 
 \begin{align*}
\mathcal{L}(w, \lambda) = w^\trsp R w - \lambda (w^\trsp w - 1).
\end{align*}
Differentiating the above with respect to \( w \) and setting the derivative to zero gives 
\[
2 R w - 2 \lambda w = 0 \implies R w = \lambda w.
\]
Thus, \( (\lambda, w) \) must be an eigenpair of \( R \). To minimize \( w^T R w \), we minimize \( \lambda w^T w \), which simplifies to minimizing \( \lambda \). Hence, the solution corresponds to the eigenvector of the smallest eigenvalue of $R$.

\section{Proof of Lemma \ref{lem:2}}\label{app:b}
Following the same steps as the proof in Appendix \ref{app:a} for Lemma \ref{lem:1}, maximizing \( w^T R w \) corresponds to selecting the maximum  eigen value \( \lambda \) of $R$. Hence, the solution is the eigenvector corresponding to  the largest eigenvalue of $R$.

\section{Proof of Lemma \ref{lem:3}} \label{app:c}
Using the transformation \( w = \epsilon_0^{-\frac{1}{2}} A^{\frac{1}{2}} z \), the minimization problem reduces to:
\[
\text{minimize} \quad \epsilon_0 w^T R w, \quad \text{subject to} \quad w^T w = 1.
\]
From Lemma \ref{lem:1}, the solution is \( w \) as the eigenvector $w'$ of the smallest eigenvalue of \( R \). Hence, the optimal solution to the original problem is given as
\[
z = \epsilon_0^{\frac{1}{2}} A^{-\frac{1}{2}} w'.
\]

Similarly, using transform \( w = l_0^{-\frac{1}{2}} B^{\frac{1}{2}} z \), the maximization converts to the problem given in Lemma \ref{lem:2}. 


\end{document}